\documentclass[twocolumn,showpacs,amsmath,amssymb]{revtex4}

\usepackage{graphicx}
\usepackage{dcolumn}
\usepackage{bm}
\newcommand{\etal}{\emph{et al.}}
\newcommand{\be}{\begin{equation}}
\newcommand{\ee}{\end{equation}}
\newcommand{\bfig}{\begin{figure}}
\newcommand{\efig}{\end{figure}}
\newcommand{\incl}{\includegraphics}

\begin{document}      

\title{Field-enhanced diamagnetism in intense magnetic field in the pseudogap state 
of the cuprate $\rm Bi_2Sr_2CaCu_2O_{8+\delta}$
} 
\author{Yayu Wang$^1$, Lu Li$^1$, M. J. Naughton$^2$, G. D. Gu$^3$, S. Uchida$^4$, N. P. Ong$^1$
}
\affiliation{
\mbox{$^1$Department of Physics, Princeton University, New Jersey 08544, U.S.A.}\\
\mbox{$^2$Department of Physics, Boston College, Chestnut Hill, Massachusetts 02467, U.S.A.}\\
\mbox{$^3$Department of Physics, Brookhaven National Laboratory, Upton, N.Y. 11973.}\\
\mbox{$^4$School of Frontier Science, University of Tokyo, Tokyo 113-8656, Japan.}
}

\date{\today}      
\begin{abstract}
In hole-doped cuprates, Nernst experiments imply that the superconducting 
state is destroyed by spontaneous creation of vortices which destroy phase coherence.  
Using torque magnetometry on $\rm Bi_2Sr_2CaCu_2O_{8+\delta}$, we uncover a field-enhanced
diamagnetic signal $M$ above the transition temperature $T_c$ that increases with applied field 
to 32 Tesla and scales just like the Nernst signal.  The magnetization results above $T_c$ distinguish $M$ from 
conventional amplitude fluctuations, and strongly support the vortex scenario for 
the loss of phase coherence at $T_c$.  
\end{abstract}

\maketitle                   
In conventional superconductors, the superconducting transition involves vanishing of the 
macroscopic wave function $\hat{\Psi}$,
but in hole-doped cuprates there is growing evidence
that the transition is caused by the proliferation of vortices which 
destroy long-range phase coherence.  
The detection of a large Nernst signal $e_N$ and kinetic inductance above the critical 
transition temperature $T_c$ has provided
evidence for the vortex scenario~\cite{Xu,WangPRB,WangPRL,WangSci,Ong,Corson},
but there should exist a magnetic signature.
Despite the loss of phase coherence above $T_c$, one should observe
a weak magnetization $M$ that differs qualitatively from `fluctuation 
diamagnetism' observed in low-$T_c$ superconductors. 
However, the magnetization evidence to date is ambiguous.
Using high-field, high-resolution torque magnetometry on
$\rm Bi_2Sr_2CaCu_2O_{8+\delta}$ (Bi 2212), we show the existence 
of a field-enhanced diamagnetism above $T_c$ that closely matches the Nernst 
signal as a function of both field $H$ and temperature $T$.  In addition to establishing the
unusual nature of the transition and its diamagnetic state above $T_c$, we show that
the upper critical field $H_{c2}(T)$ remains very large at $T_c$, a behavior
similar to that predicted for the Kosterlitz-Thouless (KT) transition~\cite{Doniach}.

\bfig[h]			
\incl[width=8cm]{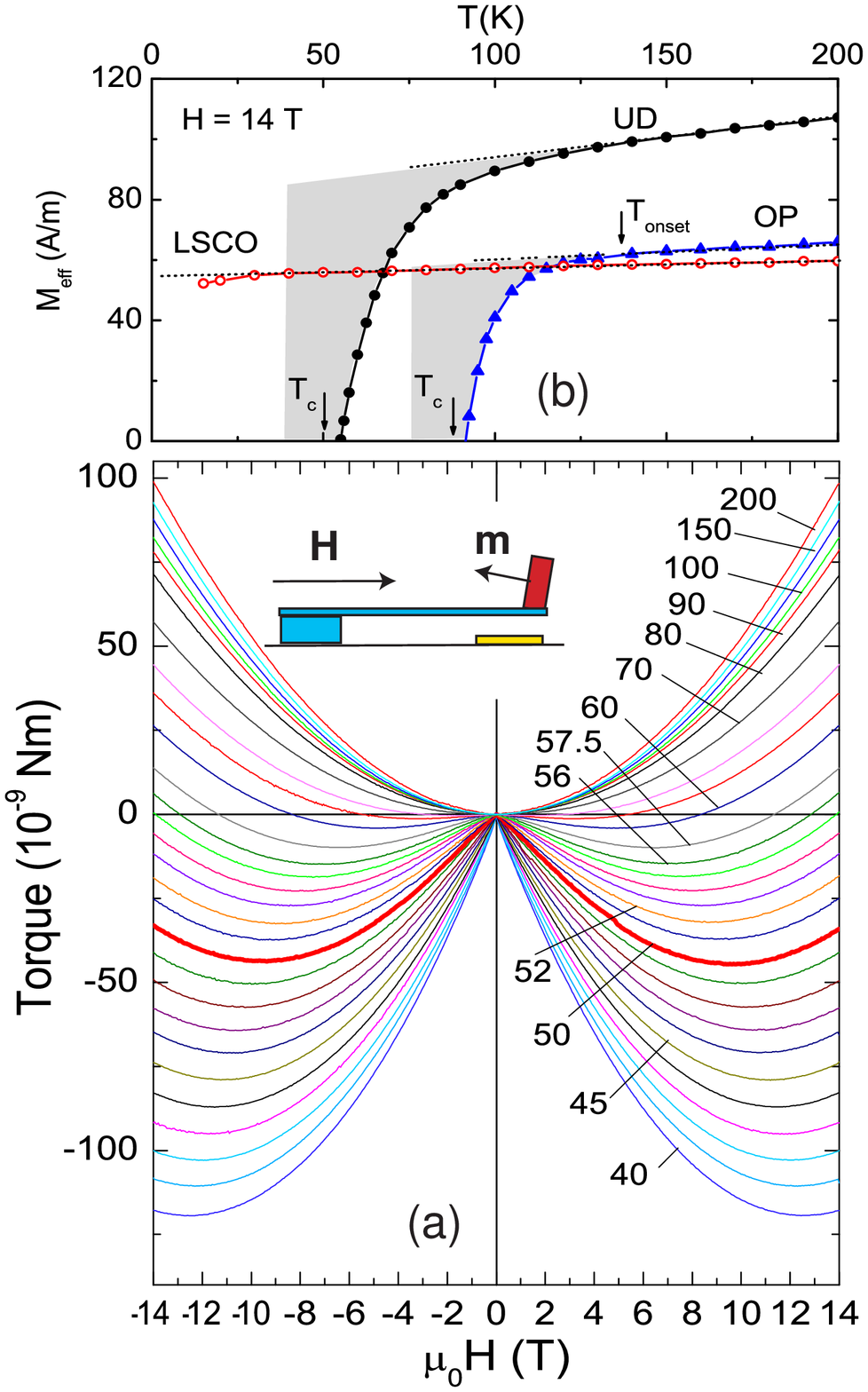}
\caption{\label{torque} (a) The measured torque $\tau$ vs. $H$ at 
selected $T$ in UD Bi 2212 with $T_c$ = 50 K 
(crystal size $\sim 0.2\times 1\times 1$ mm$^3$).  The parabolic behavior above 120 K 
arises from $\Delta\chi H_z$.  Below 120 K, a diamagnetic 
contribution $M$ grows rapidly.  Measurements were extended to 32 T at selected $T$.
The inset shows the cantilever.  The maximum beam deflection is 0.15$^{\mathrm o}$.  
(b) The $T$ dependence of $M_{eff}$ in single-crystal UD and OP Bi 2212 (solid symbols) and in 
LSCO ($x$ = 0.050, open circles) at $B$ = 14 T.  The LSCO data show 
that $\Delta\chi$ is only weakly $T$-dependent down to $\sim 25$ K.  
In Bi 2212, the diamagnetic signal $M(H_z)$ is shown shaded.
}
\efig

Torque magnetometry measurements were performed on an underdoped (UD),
an optimally-doped (OP) and an overdoped (OD) crystal of Bi 2212.  Each crystal 
was glued to the tip of a Si cantilever with 
its $c$-axis at an angle $\varphi_0\sim 15^{\mathrm o}$ 
to $\bf H$ (Fig. \ref{torque}a, inset).  The 
torque $\mbox{\boldmath{$\tau$}} = {\bf m\times B}$ leads 
to a flexing of the cantilever which is detected capacitively, where $\bf m$ 
is the sample's magnetic moment 
and ${\bf B} = \mu_0({\bf H +M})$ with $\mu_0$ the vacuum permeability.  
We resolve $m\sim 5\times 10^{-9}$ emu at 10 T.  Measurements were also performed 
in a SQUID magnetometer (with resolution $\sim 10^{-6}$ emu).
All curves of $M$ reported here are \emph{fully reversible} in $H$ and $T$.  
Our high-field results, in combination with the Nernst effect, point to conclusions 
very different from those inferred from earlier torque experiments~\cite{Bergemann,Naughton,Hofer}. 

The torque curves measured in an underdoped (UD) Bi 2212 crystal, with $T_c$ = 50 K, 
are shown in Fig. \ref{torque}a.  Above 120 K, $\tau$ is dominated by a paramagnetic term that changes
little from 200 to 120 K.  Below 120 K, however, a diamagnetic term appears, and grows 
rapidly to pull the torque to large negative values.  

We express the torque as an effective magnetization~\cite{Bergemann}
$M_{eff}\equiv\tau/B_xV$, with $V$ the sample volume and $B_x = B\sin\varphi_0$ (we take $\bf \hat{z}||\hat{c}$).
For $\varphi_0\ll 1$, we have $M_{eff} = \Delta\chi H_z + M(T,H_z)$,
where $M(T,H_z)$ is the magnetization of interest here.  The paramagnetic background
reflects the susceptibility anisotropy $\Delta\chi = \chi_c-\chi_{ab}$, which is
the difference between the uniform susceptibilities $\chi_c$ (${\bf H||\hat{c}}$) 
and $\chi_{ab}$ (${\bf H \perp\hat{c}}$).  
For either axis $i$ ($c$ or $ab$), $\chi_i$ is comprised of 3 terms, viz. 
$\chi_i(T) = \chi_{i}^{core} + \chi_{i}^{orb} + \chi_{i}^{s}(T)$~\cite{Johnston,Takigawa,Alloul}. 
The strongly anisotropic orbital (van Vleck) term $\chi_{i}^{orb}$ gives the
largest contribution to $\Delta\chi$, while the isotropic core term $\chi^{core}_i$ 
gives zero.  These 2 terms are $T$ independent, but the
spin susceptibility $\chi^{s}_i(T)$ is $T$ dependent.  
NMR (nuclear magnetic resonance) Knight-shift experiments~\cite{Takigawa,Alloul}
reveal that, in UD cuprates, $\chi^{s}_i(T)$ decreases below $T^*$, reflecting the growth of the spin gap. 
However, because the $g$-factor anisotropy is weak ($g_c/g_{ab} \sim$ 1.14), this translates to
only a small, $T$-dependent correction to the large, constant van Vleck contribution in $\Delta\chi$
(see analysis in Ref. \cite{Johnston}).

Consistent with this, our torque results reveal that the
paramagnetic term $\Delta\chi H_z$ is weakly $T$ dependent in all samples tested.
In $\rm La_{2-x}Sr_xCuO_4$ (LSCO) 
with $x$ = 0.050, in which $T_c<$2 K and $M(T)$ is not resolved 
above 25 K, $M_{eff}(T) = \Delta\chi H_z$ changes by only $\sim 6\%$ 
between 200 and 25 K (open circles in Fig. \ref{torque}b).  Likewise, 
in both Bi 2212 samples (solid symbols), $M_{eff}(T) = \Delta\chi H_z$
shows a weak $T$-dependence from 200 K to 120 K which we fit to a straight line
(dotted lines in Fig. \ref{torque}b).  We assume that $\Delta\chi$ continues this linear behavior
below $T_{onset}\sim$120 K where $M(T,H_z)$ is first resolved, and measure
$M(T,H_z)$ relative to the dotted lines (shaded areas)~\cite{background}.   
Hereafter, we write $H_z$ as $H$.

\bfig[h]			
\incl[width=8.5cm]{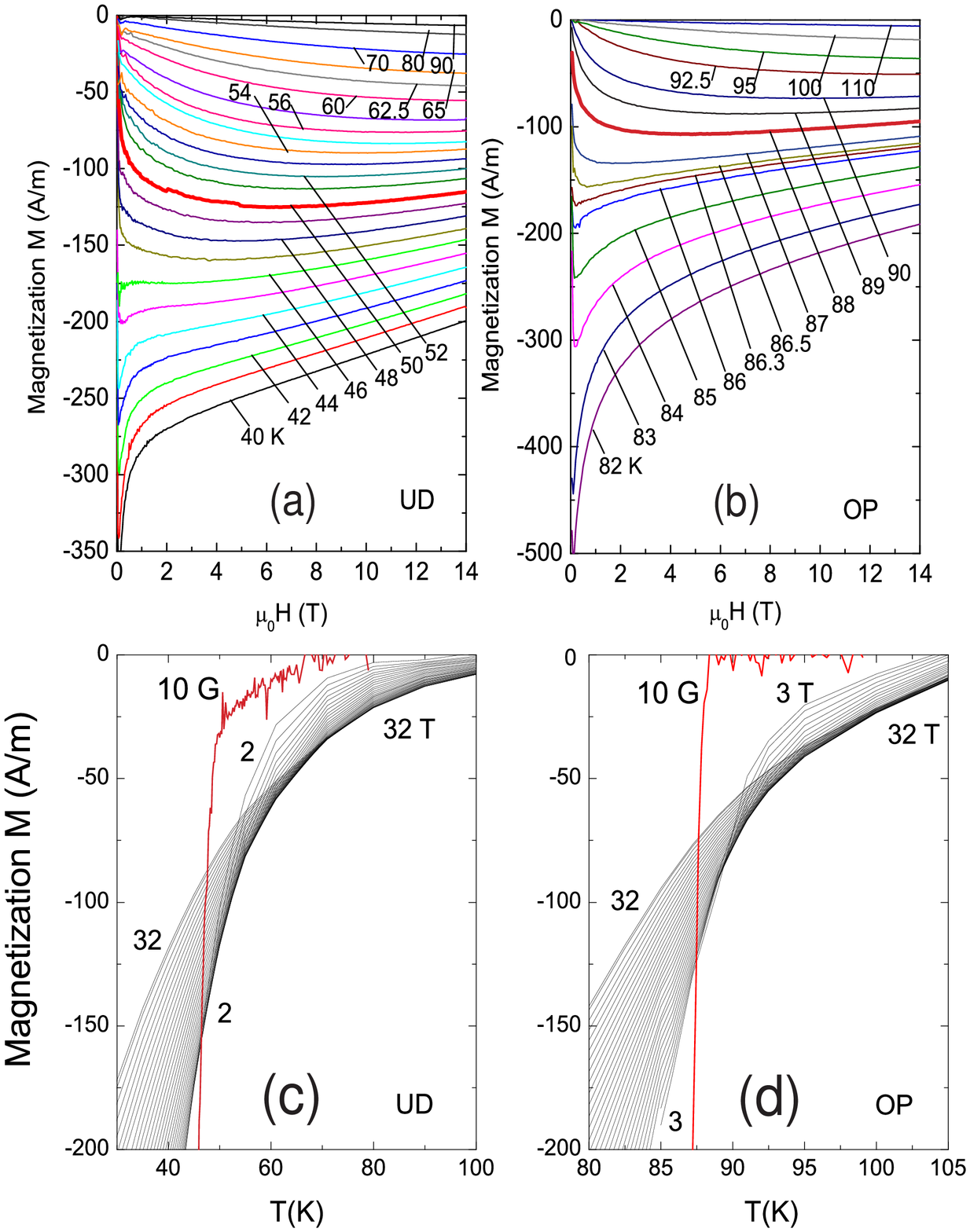}
\caption{\label{MHMT} Curves of magnetization $M(T,H)$ plotted vs. $H$ 
at selected $T$ in the (Panels a and b),  and plotted vs. $T$ in (c and d) for the UD ($T_c$ = 50 K) and 
OP ($T_c$ = 87.5 K) samples.   In (a) and (b), the bold curve is taken at $T_c$.  In Panels (c) and (d), 
the $T$ dependence of $M$ is plotted at fixed $H$ in the UD and OP sample, respectively
[$H$ decreases, in steps of 1 Tesla, from 32 T to 2 T in (a) and 3 T in (b)].  Curves labelled 10 G
show the Meissner transition at $T_c$ measured at $H$ = 10 Oe.  In the UD sample, the foot
above 50 K is a minority phase 2.5$\%$ in volume (see text).  Increasing $H$ to values 3-32 T
greatly amplifies the diamagnetic signal in a broad interval above $T_c$.
}
\efig

Figure \ref{MHMT} compares the curves of $M(T,H)$ vs. $H$ 
measured at fixed $T$ in the UD and OP samples (Panels a and b, respectively).  
As $T\rightarrow T_c$ (50 and 87.5 K, respectively), $M(T,H)$
grows over a broad interval (of width 70 K and 30 K, respectively).  
A feature of $M$, distinguishing it from the 
paramagnetic signal, is its pervasive nonlinearity versus $H$
(analyzed in Ref.~\cite{Lu}).  A second striking feature is that 
the curves 5-10 K above $T_c$ are similar in form to that
measured either at $T_c$ (shown as bold curves) or a few K below.  
To show this more clearly, we plot in Figs. \ref{MHMT}c and \ref{MHMT}d
$M$ versus $T$ with $H$ fixed at values up to 32 T.  Whereas the Meissner transition
is quite sharp (curve at 10 Oe), the high-field curves vary smoothly across $T_c$.


Previously, measurements of $M$ above $T_c$ were largely identified~\cite{Johnston,Vidal} with 
amplitude fluctuations $\delta|\hat{\Psi}|$, in analogy with ``fluctuation diamagnetism'' $M'$ in 
low-$T_{c}$ superconductors ~\cite{Gollub}.  However, difficulties
with this interpretation have been noted also~\cite{Welp,Li,Kogan,Bergemann,Naughton}.

\bfig[h]			
\incl[width=8cm]{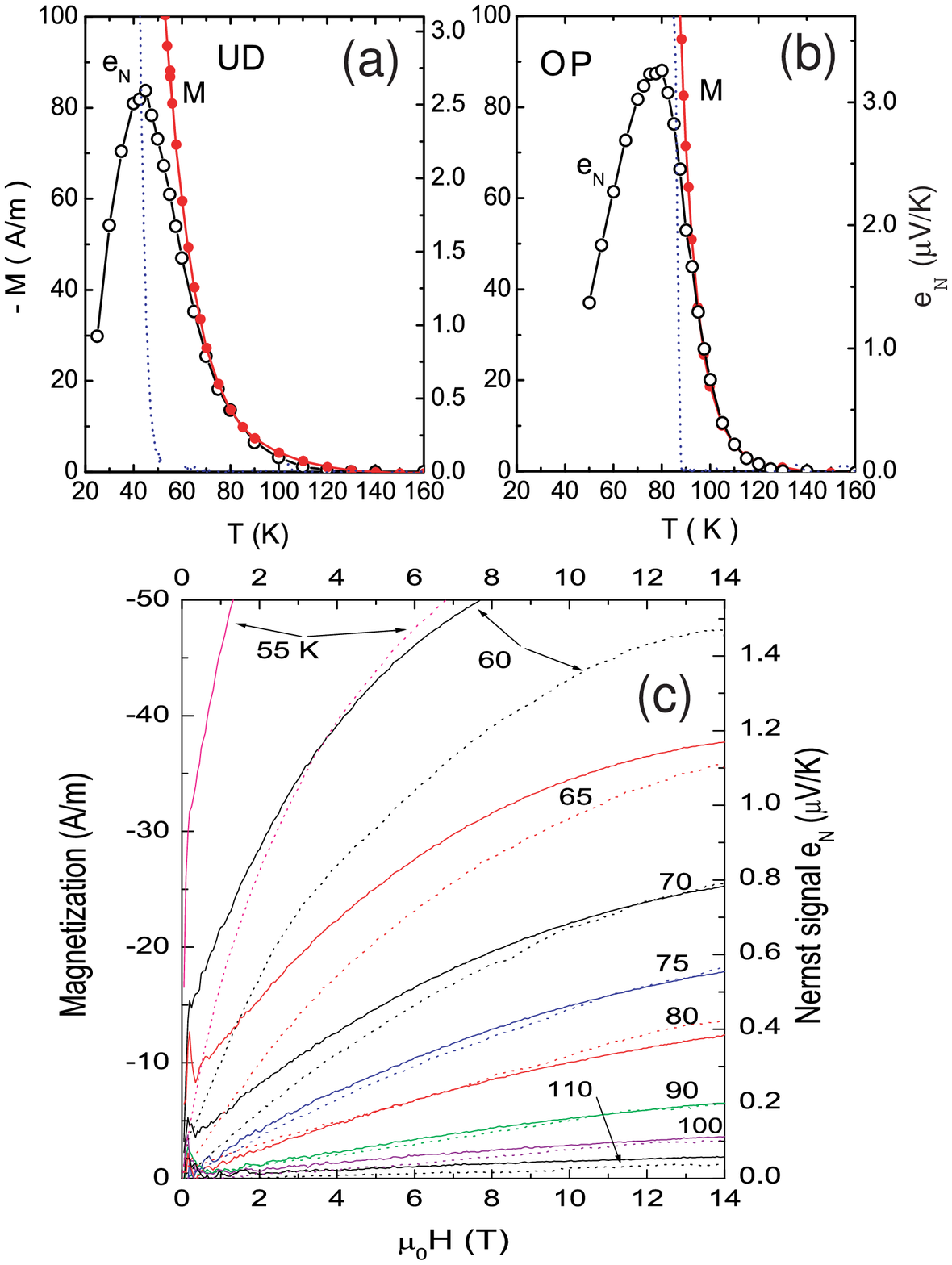}
\caption{\label{MNT} Comparison of $M$ and the vortex-Nernst signal $e_N$ 
measured at 14 T in the UD (Panel a) and OP (b) Bi 2212, and comparison of their field 
profiles in UD Bi 2212 (Panel c).  In Panels (a) and (b), $M$ and $e_N$ track each other at 
high $T$.  Below $T_c$, $M$ increases rapidly while $e_N$ 
attains a peak before falling towards zero in the vortex-solid phase.  The dashed curves 
show $M$ measured in $H$ = 10 Oe.  Panel (c) shows that 
curves of $M$ vs. $H$ (solid curves) match those of $e_N$ vs. $H$ (dashed curves) in the 
UD sample above $T_c$ (the scale factor between $M$ and $e_N$ is the same as in 
Panel a).  The observed Nernst signal is the sum of the vortex contribution
and a negative, quasiparticle (qp) term $e^{obs}_N = e_N + e^{qp}_N$.  The curves show $e_N$ after 
the small qp term is subtracted ($|e^{qp}_N| < 0.04 \mu$V/K).
}
\efig

We show next that the diamagnetism in Bi 2212 is actually qualitatively 
distinct from amplitude fluctuations.  The evidence are of 3 types.  First, we focus on $T> T_c$.  
As seen in Fig. \ref{MHMT}, in a field 
of 32 T, $M$ survives as a long tail over a 70-K interval (30-K interval) 
above $T_c$ in the UD (OP) sample.  This robustness in intense fields sharply distinguishes the cuprate 
signal from that in low-$T_c$ superconductors.  To emphasize this important
difference, we focus on the OP sample (Fig. \ref{MHMT}d).  
With $H$ = 10 Oe, $M$ displays a sharp Meissner transition at $T_c$ = 87.5 K.  However, 
a field of 3 T ``amplifies'' the diamagnetic signal by $\sim$3 orders of magnitude 
rendering $M$ observable to $\sim$104 K.  
Further increase of $H$ to 32 T makes the signal visible to 120 K.  The monotonic increase
of $M$ with $H$ implies that the condensate is not destroyed in a 32-Tesla field, i.e. the depairing field
$H_{c2}$ lies significantly higher at these $T$.  
In the UD sample (Fig. \ref{MHMT}c), a phase of volume 2.5$\%$ (estimated from $\chi$)
with higher $T_c\sim$70 K is apparent as a ``foot" extending from 50 to 70 K.  As discussed
below, this small phase does not affect our conclusions.

By contrast, in low-$T_c$ superconductors, the fluctuation signal $M'$ from
amplitude fluctuations is suppressed in weak $H$.  In Nb, $M'$ becomes unresolved above 1000 Oe 
(Fig. 13 of Ref. \cite{Gollub} for Nb, In and Pb and alloys).  
There the field sensitivity reflects the approach $H_{c2}(T)\rightarrow 0$ 
at $T_c$, and the role of nonlocal electrodynamics in suppressing 
short-wavelength fluctuations.

Secondly, we show that the diamagnetic signal above $T_c$ is closely related to the 
Nernst signal (measured in the same crystals).
In Figs. \ref{MNT}a and \ref{MNT}b, we plot the $T$ dependence of $e_N$ 
and $M$ (both at 14 T) in the UD and OP samples, respectively (the 10-Oe curves 
are shown as dashed curves).  Remarkably, $M$ (solid circles) tracks $e_N$ (open)
over a broad interval of temperature before diverging near $T_c$.  Below $T_c$, $M$ 
rises steeply whereas $e_N$ attains a broad peak before decreasing towards zero 
(its value in the vortex solid phase).  The 2 signals $M$ and $e_N$ 
share the same onset temperature $T_{onset}$.  

\bfig[h]			
\incl[width=8.5cm]{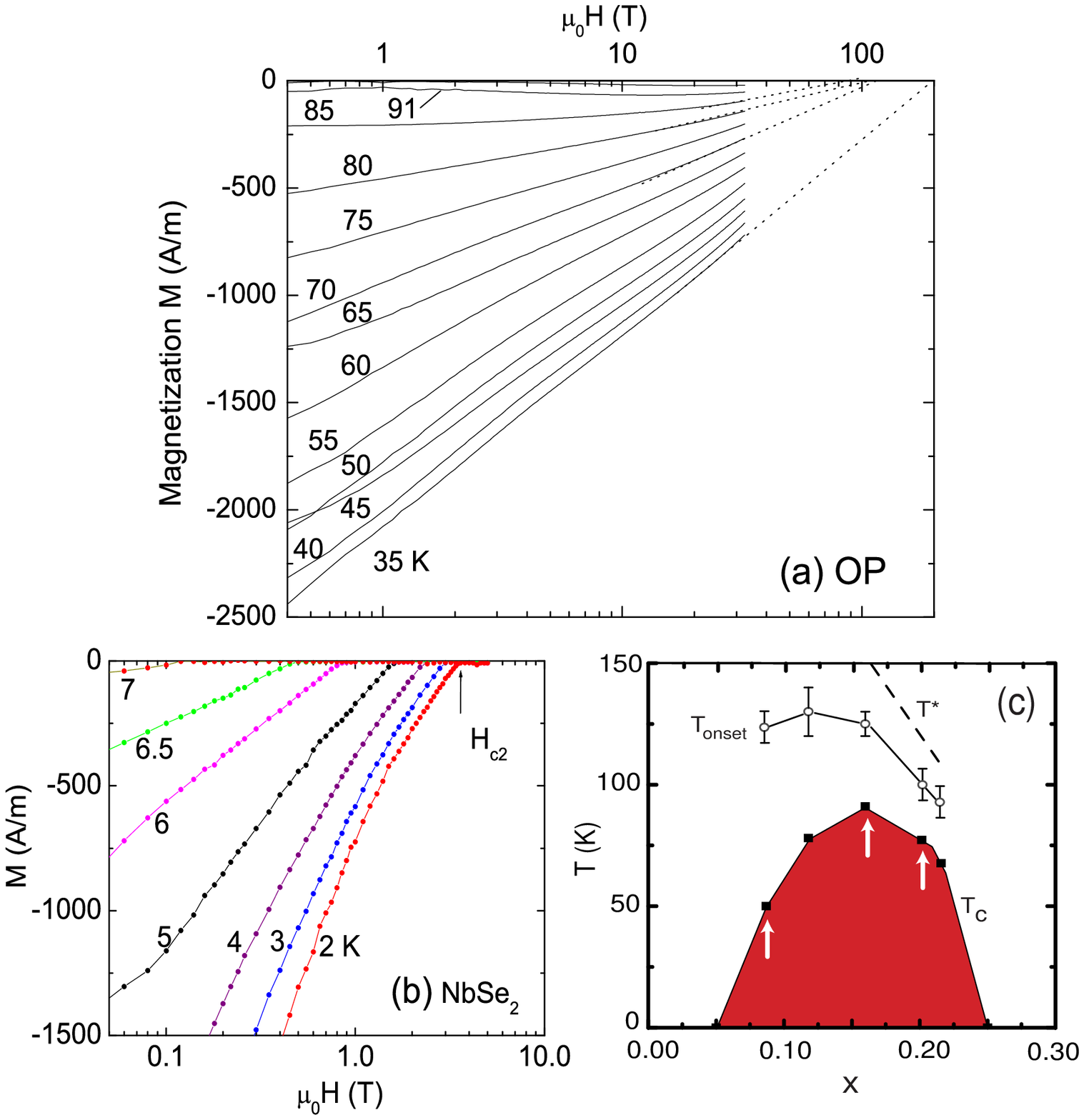}
\caption{\label{Hc2}  Plot of the field dependence of $M$ in OP Bi 2212 (Panel a) and in
$\rm NbSe_2$ (b) and the phase diagram of Bi 2212 (Panel c).  In Panel a, 
$|M|$ falls roughly linearly in $\log{H}$ from 0.2 T to 32 T.  The inferred
$H_{c2}(T)$ values stay above 90 T as $T\rightarrow T_c$
(= 86 K).  In Panel b, the measured $M$ also shows a nominally
linear dependence on $\log{H}$.  However, in contrast with (a), $H_{c2}(T)$ 
decreases from 3.6 T to 0.12 T as $T$ rises from 2 to 7 K.
In Panel c, $T_{onset}$ of both $M$ and $e_N$ is plotted vs. $x$ (hole content) 
together with $T_c$ and $T^*$.  Arrows indicate the 3 samples studied by torque magnetometry. 
}
\efig

In both samples, the direct proportionality between $M$ and $e_N$ above 
$T_c$ also holds as $H$ is varied.  We compare their respective 
field profiles in the UD sample in Fig. \ref{MNT}c.  Over the interval 70-120 K, 
we find that the $M$ vs. $H$ curves (solid) can be overlaid on the $e_N$ 
vs. $H$ curves (dashed) with the same scaling factor as in Fig. \ref{MNT}a.  
(Figure \ref{MNT}c also clarifies the contribution of the minority phase in the UD sample.  
At 60 and 65 K, the 2.5$\%$ phase adds a small term to $M$
that is nearly constant in $H$.  However, it does not contribute to $e_N$ because it does not
extend over the sample, so the curves of $M$ lie slightly above $e_N$ below 70 K.
Above 70 K, this difference is unobservable.  The 2.5$\%$ phase cannot account for the
large diamagnetic signal extending to 120 K.)
In terms of the resistivity $\rho$ and $\alpha_{yx} = J_y/|\nabla T|$, we 
have $e_N = \rho\alpha_{xy}$, with $J_y$ the transverse charge current.   
The scaling relationship~\cite{Maki} is then $\alpha_{xy} = -\beta M$ for $T>T_c$.  
Figure \ref{MNT} directly confirms that, above $T_c$, the growth of $e_N$ is accompanied by an increase 
of the diamagnetic signal, as required by the vortex scenario.

Lastly, we describe the behavior of $H_{c2}(T)$ inferred from the curves at $T<T_c$ (Fig. \ref{Hc2}a).  
In the extreme type II case, the high-field behavior $M\sim -[H_{c2}(T)-H]$, provides a 
reliable way to find $H_{c2}(T)$.  In Fig. \ref{Hc2}b we plot $M$ 
in $\mathrm{NbSe_2}$ (measured by SQUID) against $\log{H}$.  As $T\rightarrow T_c^{-}$, 
the inferred values approach zero as $H_{c2}(T)\sim (T_c-T)$.  The vanishing of $H_{c2}(T)$ 
causes the amplitude fluctuations to be sensitive to field.

The curves of $H_{c2}(T)$ in Bi 2212 behave in a qualitatively different way.  
Figure \ref{Hc2}a plots the curves of $M$ vs. $\log{H}$ in the OP sample 
from 0.4 T to 32 T at temperatures 35 to 91 K.
At each $T$, the decrease in $M$ is nominally linear in $\log{H}$, but $M$ clearly retains significant 
strength at 32 T.  Assuming this linear dependence holds above 32 T
(as the curves for $\rm NbSe_2$ suggest), we may estimate $H_{c2}$ using the
dashed lines in Fig. \ref{Hc2}a.  The inferred values of $H_{c2}$ decrease from 
200 T at 35 K to the large value 90 T at $T_c$ = 86 K, instead of
going to zero (in agreement with $H_{c2}$ derived from $e_N$~\cite{WangSci}).  
This unusual behavior of $H_{c2}(T)$ -- so strikingly different from the BCS scenario --
is similar to that predicted for the KT transition~\cite{Doniach,Vadim}.  It seems
to be a defining signature of the phase-disordering scenario.

The existence of a large $e_N$ and $M$ in an extended (``Nernst'') region above 
the $T_c$ dome has important implications for the phase diagram 
and the pseudogap state~\cite{LeeReview}.  
First, the results support the proposal~\cite{Emery} that the curve of $T_c$ vs. $x$
reflects strong phase disordering, rather than the vanishing of $\hat{\Psi}$.
Secondly, we note that the curve of $T_{onset}$ lies significantly lower 
than $T^*$ (Fig. \ref{Hc2}c), and has a different $x$ dependence.  The Nernst 
region $T_c<T<T_{onset}$ is characterized by the existence of vorticity and weak diamagnetic 
currents.  Above $T_{onset}$, however, these signatures vanish.  Hence the 
high-$T$ region $T_{onset}<T<T^*$ must harbor a type of broken-symmetry state 
in which only the spin degrees see a ``spin-gap'', but supercurrents are absent.
There are 2 distinct crossover temperatures: The local correlation that 
appears at $T^*$ affects mostly the spin degrees, whereas vorticity and supercurrents
appear at the lower $T_{onset}$.  

Finally, the $T_c$ curve is nested within the $T_{onset}$ curve, which 
is nested, in turn, within the pseudogap region below $T^*$.  This sequential nesting
suggests that, despite the absence of supercurrents, the high-temperature 
pseudogap state is closely related to $d$-wave superconductivity.  
As $T$ decreases from 300 K, the system gradually evolves 
from one to the other across the Nernst region.  Recent interesting proposals 
describe this evolution as either spin-charge locking~\cite{Anderson}, 
fluctuations of the quantization axis $\hat{I}$ in SU(2) theory~\cite{LeeReview,LeeWen}, 
or fluctuations of the ``electron nematic'' phase in the striped model~\cite{Kivelson}.

The high-field measurements were performed at the National High Magnetic Field 
Laboratory, Tallahassee, which is supported by the U.S. National Science 
Foundation (NSF) and the State of Florida.  This research is supported by 
NSF Grant DMR 0213706. GDG was supported by the DOE under 
contract No. DE-AC02-98CH10886.


\begin{thebibliography}{99}
%

\bibitem{Xu} Z. A. Xu, N. P. Ong, Y. Wang, T. Kakeshita, S. Uchida, Nature {\bf 406}, 486 (2000).

\bibitem{WangPRB} Yayu Wang \etal,  Phys. Rev. B {\bf 64}, 224519 (2001).

\bibitem{WangPRL} Yayu Wang \etal, Phys. Rev. Lett. {\bf 88}, 257003 (2002).

\bibitem{WangSci} Yayu Wang \etal, Science, {\bf 299}, 86 (2003).

\bibitem{Ong} N. P. Ong and Yayu Wang, Physica C {\bf 408}, 11-14 (2004).

\bibitem{Corson} J. Corson, R. Mallozzi, J. Orenstein, J. N. Eckstein, and I. Bozovic, Nature {\bf 398}, 221 (1999).

\bibitem{Doniach} S. Doniach and B. A. Huberman, Phys. Rev. Lett. {\bf 42}, 1169 (1979); 

\bibitem{Bergemann} C. Bergemann \etal, Phys. Rev. B {\bf 57}, 14387 (1998).

\bibitem{Naughton} M. J. Naughton, Phys. Rev. B {\bf 61}, 1605 (2000). 

\bibitem{Hofer} J. Hofer \etal, Phys. Rev. B {\bf 62}, 631 (2000).

\bibitem{Johnston}  D. C. Johnston and J. H. Cho, Phys. Rev. B {\bf 42}, 8710 (1990).

\bibitem{Takigawa} M. Takigawa \etal, Phys. Rev. B {\bf 39}, 300 (1989); \emph{ibid} Phys. Rev. B {\bf 43}
247 (1991).

\bibitem{Alloul} H. Alloul, A. Mahajan, H. Casalta, and O. Klein, Phys. Rev. Lett. {\bf 70}, 1171 (1993).

\bibitem{background} In the UD sample, a fair question
is whether the extrapolation should have a slight curvature.  We estimate that this uncertainty
adds at most an uncertainty of $\pm 7\%$ in determining $T_{onset}$.  However, because
$|M(T)|$ increases steeply below $\sim$100 K, the background uncertainty becomes increasingly 
insignificant for $|M(T)|$.
 
\bibitem{Lu} Lu Li, Yayu Wang and N. P. Ong \etal, Europhys. Lett., \emph{in press}, cond-mat/050761.

\bibitem{Vidal} C. Carballeira, J. Mosqueira, R. Revcolevschi, and F. Vidal, Phys. Rev. Lett. {\bf 84}, 3517 (2000).

\bibitem{Gollub} J. P. Gollub, M. R. Beasley, R. Callarotti, and M. Tinkham, 
Phys. Rev. B {\bf 7}, 3039 (1973).

\bibitem{Welp} U. Welp \emph{et al.}, Phys. Rev. Lett. {\bf 67}, 3180 (1991).

\bibitem{Li} Q. Li \etal, Phys. Rev. B {\bf 48}, 9877 (1993).

\bibitem{Kogan} V. G. Kogan \etal, Phys. Rev. Lett. {\bf 70}, 1870 (1993).


\bibitem{Maki} C. Caroli, and K. Maki, Phys. Rev. {\bf 164}, 591 (1967).

\bibitem{Vadim} V. Oganesyan, D. A. Huse and S. L. Sondhi, cond-mat/0502224.

\bibitem{LeeReview} P. A. Lee, N. Nagaosa, X. G. Wen, cond-mat/0410445.

\bibitem{Emery} V. J. Emery and S. A. Kivelson, Nature {\bf 374}, 434 (1995).

\bibitem{Anderson} P. W. Anderson, Phys. Rev. Lett. \emph{in press}.
 
\bibitem{LeeWen} P. A. Lee and X.-G. Wen, Phys. Rev. B {\bf 63}, 224517 (2001).

\bibitem{Kivelson} For a review, see E. W. Carlson, V. J. Emery, S. A. Kivelson, D. Orgad,
cond-mat/0206217.
%
\end{thebibliography}
\end{document}